\documentclass[lettersize,journal]{IEEEtran}

\usepackage{mathptmx} 

\usepackage{amsmath,amssymb,amsfonts}
\usepackage{algorithmic}
\usepackage{graphicx}
\usepackage{textcomp}
\usepackage{acronym}
\usepackage{comment}
\usepackage{multirow}
\usepackage{fancyhdr}
\usepackage[normalem]{ulem}
\usepackage[hyphens]{url}
\usepackage[sort,nocompress]{cite}
\usepackage[final]{microtype}
\usepackage[keeplastbox]{flushend}
\usepackage[bookmarks=true,breaklinks=true,letterpaper=true,colorlinks,citecolor=blue,linkcolor=blue,urlcolor=blue]{hyperref}

\usepackage[table]{xcolor}
\definecolor{LightCyan}{rgb}{0.88,1,1}

\usepackage{epstopdf}
\graphicspath{{figures/}}

\usepackage[caption=false,font=normalsize,labelfont=sf,textfont=sf]{subfig}
\usepackage{cite}
\usepackage{verbatim}
\usepackage{url}

\begin{document}

\title{PaST-NoC: A Packet-Switched Superconducting Temporal NoC}


\author{Darren Lyles, Patricia Gonzalez-Guerrero, Meriam Gay Bautista and George Michelogiannakis\\
Lawrence Berkeley National Laboratory, California, USA\\
Email: [dlyles,lg4er,mgbautista,mihelog]@lbl.gov} 



\maketitle

\acrodef{RL}{race logic}
\acrodef{NoC}{network on chip}
\acrodef{RSFQ}{rapid single flux quantum}
\acrodef{ERSFQ}{energy-efficient RSFQ}
\acrodef{SFQ}{single flux quantum}
\acrodef{FA}{first arrival}
\acrodef{LA}{last arrival}
\acrodef{I}{inhibit}
\acrodef{JTL}{Josephson transmission line}
\acrodef{JJ}{Josephson junction}
\acrodef{SQUID}{superconducting quantum interference device}
\acrodef{M}{merger}
\acrodef{S}{splitter}
\acrodef{DFF}{data flip flop}
\acrodef{DFF2}{two-output data flip flop}
\acrodef{UR}{uniform random}
\acrodef{WC}{worst case}
\acrodef{FF}{flip flop}
\acrodef{TFF}{toggle flip flop}
\acrodef{NDRO}{non destructive read out}
\acrodefplural{NoC}[NoCs]{networks on chip}
\acrodef{SRNoC}{superconducting rotary NoC}
\acrodef{PaST-NoC}{packet-switched superconducting temporal NoC}
\acrodef{HPC}{high performance computing}
\acrodef{PRNG}{pseudo random number generator}
\acrodef{LFSR}{linear-feedback shift register}
\acrodef{EDA}{electronic design automation}
\acrodef{PE}{processing element}
\acrodef{VC}{virtual channel}
\acrodef{SQUID}{superconducting quantum interference device}
\acrodef{DOR}{dimension order routing}

\begin{abstract}

Temporal computing promises to mitigate the stringent area constraints and clock distribution overheads of traditional superconducting digital computing. To design a scalable, area- and power-efficient superconducting \ac{NoC}, we propose \ac{PaST-NoC}. \ac{PaST-NoC} operates its control path in the temporal domain using \ac{RL}, combined with bufferless deflection flow control to minimize area. Packets encode their destination using \ac{RL} and carry a collection of data pulses that the receiver can interpret as pulse trains, \ac{RL}, serialized binary, or other formats. We demonstrate how to scale up \ac{PaST-NoC} to arbitrary topologies based on 2$\times$2 routers and 4$\times$4 butterflies as building blocks. As we show, if data pulses are interpreted using \ac{RL}, \ac{PaST-NoC} outperforms state-of-the-art superconducting binary \acp{NoC} in throughput per area by as much as $5\times$ for long packets.

\end{abstract}


\acresetall 

\section{Introduction}
\label{section:introduction}

\IEEEPARstart{S}{uperconducting} digital computing is a promising alternative to CMOS due to its ability to operate at several tens of GHz at a higher energy efficiency than CMOS~\cite{Superconducting_accelerators,RSFQ}.
However, the full potential of superconducting digital computing is hindered by three major factors: First, the majority of superconducting digital \ac{RSFQ}~\footnote{RSFQ is currently the most developed superconducting logic family.} gates such as AND, OR, and XOR are synchronous, increasing the clock tree overhead and making scaling up challenging due to the need for precise timing~\cite{SFQ_clock,Asynchronous_gates,Race_logic_rsfq}. Second, superconducting technology suffers from limited device density~\cite{RSFQ_scalability}. In fact, recent \ac{RSFQ} chips were restricted to just a few tens of thousands of \acp{JJ}, the fundamental switching device in superconducting computing. This limited density is exacerbated by the need of splitters which are interconnection cells that implement fanout, because of the severely limited inherent fanout of gates~\cite{SFQ_clock,Asynchronous_gates,Race_logic_rsfq}. Third, superconducting memory is particularly area expensive~\cite{zokaee_2021_micro}. These limitations, combined with adopting large, CMOS-inspired \ac{NoC} control circuits and complicated wiring for wide buses to represent packets in binary, reduce area efficiency and restrict recent binary \ac{RSFQ} \acp{NoC} to eight inputs and eight outputs at most~\cite{Crossbar3,crossbar_switch_4x4,Crossbar2,Banyan2,Banyan_network,Ring,Switch_scheduler}.

\begin{figure}
\centering
  \includegraphics[width=\columnwidth]{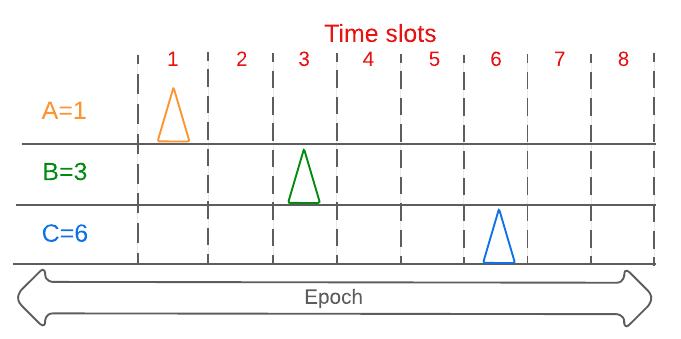}
  \caption{\Acf{RL} encodes values as time of arrival of pulses. Time is divided into repeating epochs that are divided into time slots, each denoting a value. A pulse arriving in time slot ``I'' represents the value ``I''.\label{figure:rl example}}
\end{figure}

In this work, we propose \ac{PaST-NoC} that addresses current superconducting area density limitations by maximizing throughput per unit area. The key for the area efficiency of \ac{PaST-NoC} is mapping control information such as packet destination as well as operating router control paths according to the time of arrival of \ac{RSFQ} pulses using \ac{RL}~\cite{Race_logic_rsfq}, instead of binary representation. In \ac{RL}, information is mapped to the time of arrival of an \ac{RSFQ} pulse (\figurename~\ref{figure:rl example}).

\ac{PaST-NoC} is the first packet-switched \ac{NoC} in \ac{RSFQ} that operates its control path entirely in the temporal domain using \ac{RL} and uses bufferless deflection flow control to avoid the elevated area cost of superconducting memory. As such, \ac{PaST-NoC} provides tradeoffs that are more favorable to the unique characteristics of superconducting technology than traditional CMOS~\cite{Case_bufferless,NoCs_bufferless}. The additional dynamic power due to deflections has a minimal impact in \ac{RSFQ}~\cite{Energy_and_family_discussion,Superconducting_accelerators}. In addition, we propose a round-robin deflection policy~\cite{Mad_postman} with randomization to provide probabilistic livelock freedom~\cite{Chaos_router}.
These novel contributions are described in Sections~\ref{section:packet format} and~\ref{section:router architecture}.

Although \ac{PaST-NoC}'s control path operates entirely in \ac{RL} for superior area efficiency, data can be encoded as serialized binary or unary, or a different encoding. Examples of unary encoding are \ac{RSFQ} pulse streams, \ac{RL}, or stochastic streams~\cite{USFQ}. This makes \ac{PaST-NoC} a candidate for conventional architectures that use typical binary encoding, but also for non-conventional architectures that use alternative data representations. These non-conventional architectures have demonstrated the potential for integrating hundreds of \acp{PE} on today's superconducting chips~\cite{USFQ,FFT_newcas,Quantum_control_temporal,SpikingNNs,Qubit_control,RSFQ_neurons, RSFQ_neurons2}. Not only do these scales motivate the use of a \ac{NoC} but in fact previously-demonstrated binary superconducting \acp{NoC} are far from reaching the scales of hundreds that these emerging architectures call for.
In addition, temporal \acp{PE} can readily generate control information such as packet destinations in the temporal domain that naturally fits \ac{PaST-NoC}, instead of converting to binary in order to use binary \acp{NoC}.

\ac{PaST-NoC} uses 2$\times$2 routers as a building block to reduce the area and complexity of its \ac{RL}-based routing logic. Based on these 2$\times$2 routers, we describe how \ac{PaST-NoC} easily scales up to a 4$\times$4 butterfly and an 8$\times$8 mesh that form the basis for a variety of larger topologies.
To achieve this, we extend previous methods as described in Section~\ref{section:scaling up} in order to effectively use them for \ac{PaST-NoC}.
When the data portion of packets is encoded using \ac{RL}, \ac{PaST-NoC} provides a higher throughput per port per area (calculated as number of \acp{JJ}) compared to state-of-the-art binary \ac{RSFQ} \acp{NoC} by over 5$\times$, and a statically-scheduled \ac{RSFQ} temporal \ac{NoC}~\cite{SRNoC} by as much as 2$\times$.

\acresetall 
\section{Background and Motivation}
\label{section:background and motivation}

\begin{footnotesize}
\begin{table*}
\centering
\caption{A brief summary of superconducting logic primitives that appear in our study.\label{table:primitives_summary}}
\begin{tabular}{|c|c|c|c|c|}
\hline
\rowcolor{LightCyan}
Name & Ins & Outs & Summary & Num of JJs\\
\hline
\hline
\multirow{2}{*}{Splitter (S)} & \multirow{2}{*}{1} & \multirow{2}{*}{2} & Produces a pulse at both outputs when receiving & \multirow{2}{*}{3}\\
& & & an input pulse. Implements fanout. & \\
\hline
\multirow{2}{*}{Merger (M)} & \multirow{2}{*}{2} & \multirow{2}{*}{1} & Produces an output pulse when receiving & \multirow{2}{*}{5}\\
& & & a pulse at either input. Implements fanin. & \\
\hline
Last arrival (LA) & \multirow{2}{*}{2} & \multirow{2}{*}{1} & Produces an output pulse as soon as it has & \multirow{2}{*}{6}\\
A re-purposed coincidence gate~\cite{Race_logic_rsfq}&&& observed a pulse from each of its two inputs. &\\
\hline
Inhibit (I) & \multirow{2}{*}{2} & \multirow{2}{*}{1} & Propagates a pulse from input 1 unless a & \multirow{2}{*}{8}\\
A re-purposed inverter cell~\cite{Race_logic_rsfq}& & & pulse arrived at input 2 more recently than input 1. &\\
\hline
\multirow{2}{*}{Non-destructive read out (NDRO) cell~\cite{Small_NDRO}} & \multirow{2}{*}{3} & \multirow{2}{*}{1} & Input pulses propagate to output if NDRO has observed a pulse in its& \multirow{2}{*}{7}\\
&&& SET control input more recently than a pulse in its reset control input. &\\
\hline
\multirow{2}{*}{Logical AND gate (AND)} & \multirow{2}{*}{3} & \multirow{2}{*}{1} & Pulse at clock input causes output pulse if both & \multirow{2}{*}{11}\\
&&& data inputs received pulse since previous clock pulse. &\\
\hline
\multirow{2}{*}{Toggle flip flop (TFF) or (T)} & \multirow{2}{*}{1} & \multirow{2}{*}{2} & First input pulse causes pulse at output 1, & \multirow{2}{*}{10}\\
&&&second at output 2, third at output 1, and so on. &\\
\hline
Data flip flop (DFF) & 2 & 1 & Propagates input to output upon a readout (i.e.., clock) pulse. & 4\\
\hline
Two-output data flip flop (DFF2) & 3 & 2 & DFF with one data input but two readout inputs, each with its own output.& 12\\
\hline
\end{tabular}
\end{table*}
\end{footnotesize}

\subsection{Superconducting Computing}
\label{section:superconducting computing}

\subsubsection{Fundamentals}
\label{section:fundamentals}

Superconductivity is the property of certain metals to have zero resistance below a critical temperature that is usually a few Kelvins~\cite{Superconducting_accelerators}. The fundamental superconducting switching device is the \ac{JJ}~\cite{JJ}. The \ac{JJ} allows current to pass through its two terminals with no resistance until a critical current $I_c$ is reached. Reaching $I_c$ causes the \ac{JJ} to switch to a resistive state, causing a magnetic quantum flux transfer. This transfer is observable at the \ac{JJ} terminal as a voltage pulse that lasts only a few picoseconds and has an amplitude of a few mVs~\cite{RSFQ}.
Because \acp{JJ} can switch in just a few picoseconds, they enable clock frequencies of several tens of GHz~\cite{Crossbar3,Superconducting_accelerators,Race_logic_rsfq}. Moreover, superconducting digital computing promises energy-efficient computation even after
accounting for cooling, due to the low \ac{JJ} energy consumption which is several orders of magnitude lower than CMOS~\cite{Superconducting_accelerators}.
Cooling is already present in cryo-cooled systems such as quantum computers~\cite{Qubit_control} and outer space electronics~\cite{krylov2022single}.
However, there is currently an approximate 1000$\times$ area density gap compared to CMOS~\cite{Superconducting_accelerators, RSFQ_scalability}.
This limits superconducting digital circuits to small scales and memory sizes.

The most mature superconducting digital logic family is \ac{RSFQ} and its variants~\cite{RSFQ}. In
\ac{RSFQ}, a logical 1 is encoded by the presence of a pulse and a logical 0 by the absence of a pulse. Because pulses (or their absence) are short-lived and travel at the speed of light, they are unlikely to arrive at logic gates at the same time. Therefore, \ac{RSFQ} typically requires almost each gate in the design to be clocked, resulting in deeply-pipelined architectures; 
this increases the cost overhead associated with clock tree generation and distribution as well as makes timing closure challenging~\cite{SFQ_clock}. Importantly, fanout in \ac{RSFQ} requires splitter cells that use three \acp{JJ} each and have one input and two outputs.
For instance, clocking 1024 synchronous gates requires 3072 \acp{JJ} for 1024 splitters~\cite{Clock_tree_rsfq}, a significant percentage of the overall design.
\ac{RSFQ} offers a rich set of traditional gates, such as ANDs, ORs, and \acp{DFF}.
Most gates are based on a superconducting loop that includes at least one \ac{JJ}, called a \ac{SQUID}. \acp{SQUID} preserve current where the intensity and direction of flow can be used to store state.

\subsubsection{Race Logic}
\label{section:race logic}

\Acf{RL} in superconducting circuits is a form of temporal data representation and mitigates \ac{RSFQ}'s low device density by trading area with time since a single wire can encode multiple values (\figurename~\ref{figure:rl example})~\cite{Race_logic_rsfq}.
In \ac{RL}, a number is encoded as the time of arrival of a pulse within the computing time (epoch). For instance, if an epoch
lasts 400ps and we define a ``time slot'' to be 50ps, there are eight time slots and thus a pulse can encode one of eight values.
A pulse that arrives in the sixth time slot encodes the value 6, assuming the first time slot encodes the value 1.
After the last time slot, a reset that is often called an epoch signal arrives and a new epoch begins.

\ac{RL} computes by comparing the relative time of arrival of pulses. For instance, the \ac{LA} gate produces an output pulse when the second of its two inputs observes a pulse. Because RL encodes data in the temporal domain, it is well-suited for dynamic programming~\cite{Race_logic}. Compute gates in \ac{RSFQ} \ac{RL} are stateful in order to remember what pulses arrived and their relative timing.

Table~\ref{table:primitives_summary} summarizes the superconducting cells that appear in our study. We implement these cells based on designs from~\cite{Race_logic_rsfq} and SUNY's online library, except for the \acs{NDRO} that is based on~\cite{Small_NDRO}.

\subsection{NoCs and Deflection Flow Control}
\label{section:nocs and deflection flow control}

On-chip networks, also known as \acp{NoC}, typically consist of a set of routers that connect to each other and to endpoints in a manner defined by the topology~\cite{Dally_packets}.
\acp{NoC} have been commercially adopted~\cite{Intel_noc,Intel_noc2} and may use deterministic, oblivious, or adaptive routing.
Moreover, the flow control defines how packets progress through the network. Because reducing area in CMOS is usually not the top priority, CMOS \acp{NoC} typically rely on input-queued buffered routers with credit-based backpressure to prevent buffer overflow.
In contrast, using deflection flow control, a packet or flit never stalls or waits at routers~\cite{Mad_postman,Case_bufferless,NoCs_bufferless}. Instead, if contention arises by having more than one packet requesting the same output, the router picks a winning packet to proceed. The losing packets are sent to any available outputs, which may lead the packets farther away from their final destinations.
Bufferless flow control is attractive for an \ac{RSFQ} \ac{NoC} due to the stringent area constraints and the elevated cost of superconducting memory~\cite{zokaee_2021_micro}. By using deflection flow control, we eliminate router buffers at the expense of a marginal increment in dynamic power. This increment in dynamic power has a negligible impact due to \ac{RSFQ}'s dynamic energy efficiency~\cite{Energy_and_family_discussion,Superconducting_accelerators}.

\subsection{Related Work}
\label{section:related work}

In \ac{RSFQ}, demonstrated \acp{NoC} have focused on a binary representation and have been limited to no more than eight inputs and outputs because they mostly use CMOS-inspired control such as for routing and allocation, as well as multiple parallel wires to represent packets in binary. As a result, these \acp{NoC} are not adequately area-efficient to scale up further given \ac{RSFQ}'s device density~\cite{Crossbar3,crossbar_switch_4x4,Crossbar2,Banyan2,Banyan_network,Ring,Switch_scheduler}. In particular, previous work demonstrated 2$\times$2~\cite{Crossbar2, Banyan2} and $4\times$4~\cite{Crossbar3,crossbar_switch_4x4} crossbar switches including schedulers~\cite{Switch_scheduler}, a 3$\times$3 ring~\cite{Ring}, and larger mesh or Banyan networks (that are topologically equivalent to a butterfly network)~\cite{Banyan_network}.

Further past work on \acp{NoC} used temporal encoding only in the data path (not in the control path) by temporally encoding packet payloads in CMOS~\cite{Temporal_NoC} and in \ac{RSFQ}~\cite{SRNoC}. In particular, \ac{SRNoC}~\cite{SRNoC} follows a rigid rotating schedule. Thus, its performance suffers under unbalanced traffic patterns. Also, depending on the number of routers, \ac{SRNoC} may require potentially deep buffers at network boundaries for packets to wait for their desired connection to be established. Our proposed \ac{PaST-NoC} is the first packet-switched \ac{NoC} that operates its control in the temporal domain using \ac{RL} and offers a versatile data path.

\section{Packet format}
\label{section:packet format}

This and the next two sections describe the novel contributions of this paper.
Consistent with our design goal to operate \ac{PaST-NoC}'s control path entirely temporally, \ac{PaST-NoC} uses a packet format that is encoded in the temporal domain using \ac{RL}. Following the \ac{RL} convention, a packet's duration is equal to an epoch, which is defined by a periodic signal. The epoch signal serves to reset the values that \ac{RL} pulses represent and also helps re-synchronization at pre-defined \ac{NoC} boundaries, as we discuss later. The epoch duration is constant throughout the \ac{NoC}.

\begin{figure}
\centering
  \includegraphics[width=\columnwidth]{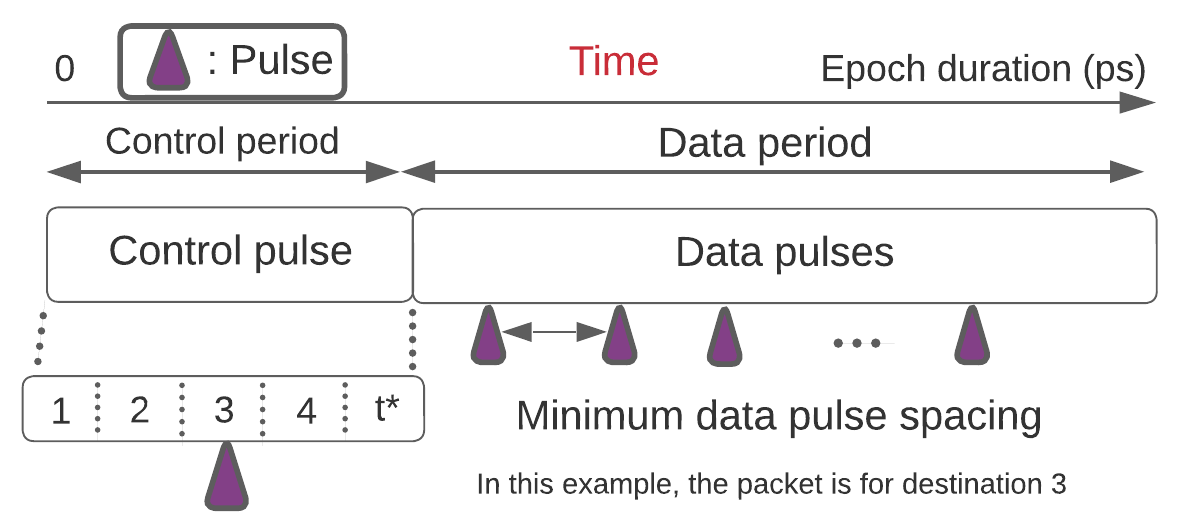}
  \caption{\acs{PaST-NoC}'s packet format. The control period has exactly one pulse encoded in \acs{RL} to signify the packet's final destination. In this example, there are four potential destinations. The last control time slot (labelled $t*$) does not contain a pulse but gives the routing logic sufficient time to produce results. Time elapses from left to right.\label{figure:packet format}}
\end{figure}

Packets traverse the \ac{NoC} as a unit and cannot be divided or merged with other packets.
As shown in \figurename~\ref{figure:packet format}, packets have control information that is succeeded by data. Control information is encoded using \ac{RL}. Control and data boundaries are defined by a duration that is a subset of an epoch that we name a ``period''. For example, if the epoch lasts 500ps, the control period can last 200ps and the data 300ps.

\subsection{Data Period}
\label{section:data period}

The data period contains pulses that are interpreted only by the receiver, not routers. Therefore, data pulses can be encoded in \ac{RL} or any other format the receiver expects. For example, if terminals are neurons in \ac{RSFQ} and \ac{PaST-NoC} implements the connection in a neuromorphic processor~\cite{RSFQ_neurons}, packets can carry pulse trains. Alternatively, packets can carry binary values that are transported serially.

Without loss of generality, in our work, we assume data pulses are encoded in \ac{RL} similar to control pulses.
To that end, data periods are divided into ``time slots'' where each slot denotes a value. For instance, a pulse in the third slot encodes the value of 3, assuming that the first slot encodes the value of 1 ($B$ in \figurename~\ref{figure:rl example}).
To compare against binary, the length of a data period determines the equivalent bit resolution that each data pulse represents. For $b$ bits of resolution, we need $2^b$ time slots. For example, with a bit resolution of 3, we can represent 8 different numbers (e.g., 0-7) and thus the packet's data period has eight time slots. The duration of the data period and the number of data pulses in each data period are application-dependent.

With this encoding, a packet can carry as many \ac{RL} pulses as time slots, as long as each time slot has at most one \ac{RL} pulse. In case of conflicts, that is when we need to send two pulses with the same \ac{RL} value, we can distribute these conflicting data pulses in different packets. Therefore, the number of data pulses per packet (the amount of data a packet carries) depends on the data's value.
We do not tailor our evaluation to a particular application. Thus, we evaluate a range of data period durations and assume that data pulses can fall at any time slot in the data period with a uniform random probability. This way, if we assume that $n$ is the number of available time slots in the packet's data period and we also try to fit $n$ data pulses per packet, we can use the bins and balls model~\cite{SRNoC} to calculate the approximate number of data pulses $m$ we can fit per packet without conflicts (at most one pulse per time slot) as:  

\begin{equation} \label{equation:equal}
    m = n - \frac{n}{e}   
 \end{equation}

where $e$ is the exponential constant or Euler's number. For example, for $b=5$ and $n=2^b=64$, the expected number of data pulses per packet is $m=40.1$.

\subsection{Control Period}
\label{section:control period}

The control period contains a single pulse that follows the \ac{RL} convention and encodes the packet's final destination. In the example of \figurename~\ref{figure:packet format}, the packet is destined to destination number three out of four possible destinations. As \ac{PaST-NoC} scales up, the control period is prolonged to contain a time slot for each potential destination. In addition,
we add a time slot at the end of the control period where we do not expect a pulse. This slot provides the routing logic sufficient time to produce a result before the end of the control period and thus configure the router's data path without violating cell setup times due to data pulses arriving prematurely. Thus, a control period for a four-destination \ac{PaST-NoC} has five time slots.
\section{Router architecture}
\label{section:router architecture}

\ac{PaST-NoC} implements bufferless deflection flow control~\cite{Mad_postman,Case_bufferless,NoCs_bufferless} to avoid the area cost of memory because \ac{RSFQ}'s restricted device density makes area a primary constraint. Moreover, the additional dynamic energy for deflections has a relatively small impact in \ac{RSFQ}. To keep design complexity low, \ac{PaST-NoC} routers have two inputs and two outputs (2$\times$2). Therefore, if two incoming packets request the same output (a conflict), one proceeds to its requested output and the other is deflected to the other output that is guaranteed to be available since there is an equal number of inputs and outputs.

\subsection{Router Overview}
\label{section:router diagram}

\begin{figure}
\centering
  \includegraphics[width=\columnwidth]{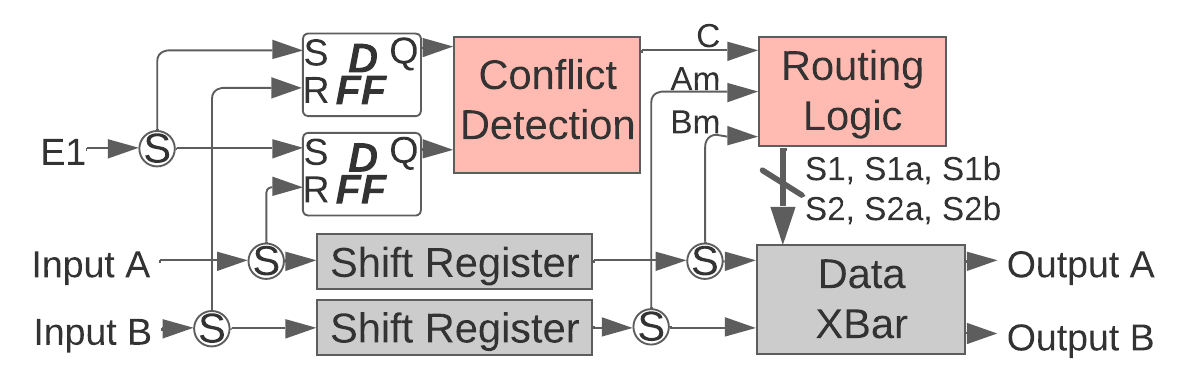}
  \caption{A \ac{PaST-NoC} 2$\times$2 router. Data path components are in gray (bottom) and control path components in red (top). ``S'' circles are splitter cells.\label{figure:router diagram}}
\end{figure}

\figurename~\ref{figure:router diagram} shows a block-level diagram of a \ac{PaST-NoC} 2$\times$2 router. The splitter and \ac{DFF} at each input (A and B) send only the first pulse that arrives at each input, i.e., the pulse at each packet's control period, to the conflict detection logic. In parallel, the entirety of the packet from each input traverses a shift register. The shift register along that path delays the packet by a period of time equal to one control period duration. This allows the conflict detection logic to produce its output $C$, which is propagated to the routing logic (Section~\ref{section:routing logic}).

The shift register has a number of stages equal to the control period divided by the minimum spacing between data pulses, since we find that it is smaller than the minimum spacing between control pulses and data pulses also traverse the shift register. Each stage has a duration equal to the minimum spacing between data pulses. This duration also defines the period of the local clock signal that each shift register requires. For instance, for a control period duration of 150ps and a data pulse spacing of 15ps, each shift register has $\frac{150}{15} = 10$ stages of 15ps each.
To reduce the \ac{JJ} count as delay scales, we base our implementation on a flux-based shift register~\cite{Flux_shift_register}.

The first pulse that exits the shift register in an epoch, which is the control pulse of each packet, is propagated to the routing logic. After receiving $C$ from the conflict detection logic, the routing logic produces its outputs. We label the delayed (post shift register) version of $A$ as $A_{m}$ and likewise $B_{m}$ for $B$. The routing logic also receives data pulses after the control pulse through $A_{m}$ and $B_{m}$, but data pulses have no effect in this block.

\ac{PaST-NoC} routers receive a threshold input ($Thr$) to help map packet destinations to router outputs; if a packet's control pulse arrives before $Thr$, it requests the first (top) router output, otherwise the second (bottom) output.
As we later explain, the time interval between epoch start ($E1$) and $Thr$ that a router receives depends on the router's position in the \ac{NoC}.

The primary outputs of the routing logic are $S1$ and $S2$. $S1$ causes the data crossbar to connect input $A$ to output $A$ and input $B$ to output $B$; $S2$ causes the opposite.
A pulse in $S1$ is asserted when there is no conflict and at least one packet from input $A$ requested output $A$ or a packet from input $B$ requested output $B$. Likewise, a pulse in $S2$ is asserted when at least one packet from input $A$ requested output $B$ or a packet from input $B$ requested output $A$. In the case of conflict, the routing logic can resolve it with a pulse at $S1$ or $S2$ because either will route one of the two packets to its desired output. Finally, if there is only one incoming packet, control signals $S1$ or $S2$ connect the input that does not receive a packet with an available output, but this has no adverse effect since no pulses traverse that path.

\subsection{Routing logic}
\label{section:routing logic}

The router's control logic is based on \ac{RL} and is responsible for decoding the desired destination of each packet, resolving conflicts, and configuring the data crossbar before data pulses traverse the crossbar. This logic can be designed to follow a fixed-priority or a round-robin scheme.
We use fixed priority as a building block for round robin. 

\subsubsection{Fixed Priority Routing logic}
\label{section:fixed priority routing logic}

\begin{figure}
\centering
  \includegraphics[width=\columnwidth]{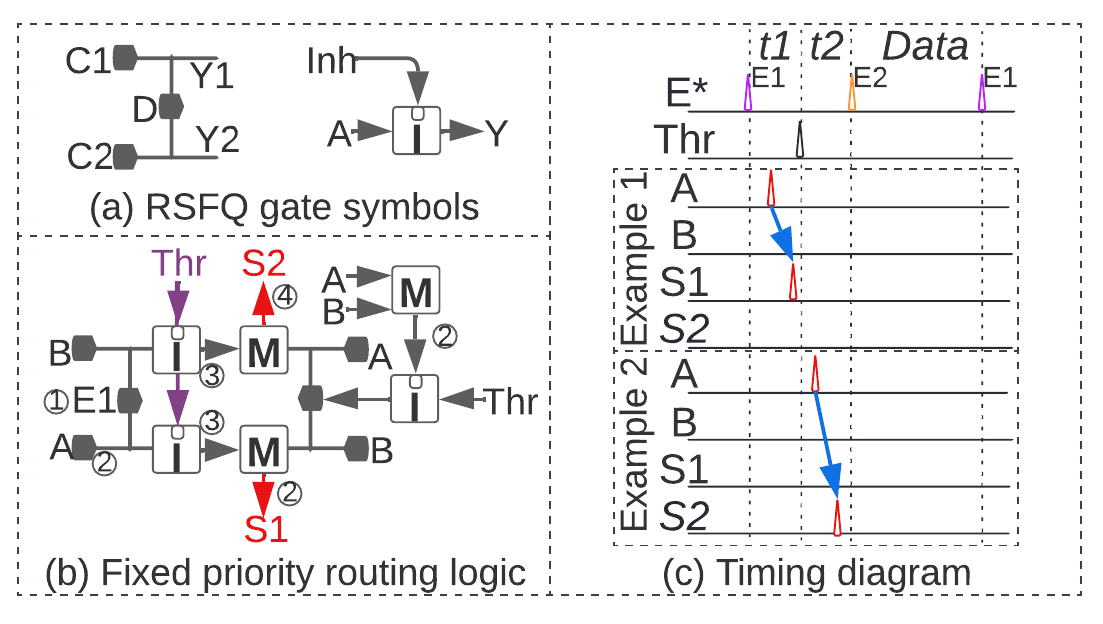}
  \caption{(a) Here we show the diagrams of two separate gates that appear later. On the left, we show a DFF2 that has two readout (or clock) inputs ($C1$ and $C2$), one data input ($D$), and two data outputs ($Y1$ and $Y2$). A pulse at input $C1$($C2$) generates a pulse at output $Y1$($Y2$) if there was a prior pulse at input $D$. On the right is an inhibit cell that has a data input ($A$), data output ($Y$), and an inhibit control input ($Inh$). An incoming pulse to input $A$ is propagated to output $Y$ unless a pulse arrived at $Inh$ more recently than the pulse at $A$.  (b) Circuit for the fixed priority routing logic. (c) An example of one packet arriving in $A$, before and after the threshold ($Thr$). $E*$ is not a separate signal but rather a notation to illustrate the arrival of both $E1$ and $E2$ in the same waveform line.\label{figure:fixed priority routing logic}}
\end{figure}

The fixed priority routing logic always resolves a conflict in favor of the packet whose control pulse arrives first. This can result in unfair bandwidth allocation, especially under heavy load. On the other hand, this compact routing logic uses a low number of \acp{JJ}. The fixed priority routing logic can be built using two \acp{DFF2} connected facing each other as shown in \figurename~\ref{figure:fixed priority routing logic}(b). The \ac{DFF2} on the left is set by pulse $E1$ that marks the beginning of the epoch \textcircled{1}. The \ac{DFF2} on the left then waits for $A$ and $B$ to arrive before $Thr$. If $A$ ($B$) arrives first, $S1$ ($S2$) is produced and the \ac{DFF2} on the right is never enabled (i.e., its data input never receives a pulse) because of the inhibit cell \textcircled{2}. In the case that neither $A$ nor $B$ arrive before the threshold ($Thr$), the \ac{DFF2} path on the left side is disabled by $Thr$ through the inhibit cells \textcircled{3} and the \ac{DFF2} on the right is enabled by $Thr$. In this case, if $A$ ($B$) arrives first, $S2$ ($S1$) is set \textcircled{4}.
\figurename~\ref{figure:fixed priority routing logic}(c) shows two examples where $S1$ and $S2$ are generated.

\subsubsection{Conflict Detection}
\label{section:conflict detection}

\begin{figure}
\centering
  \includegraphics[width=\columnwidth]{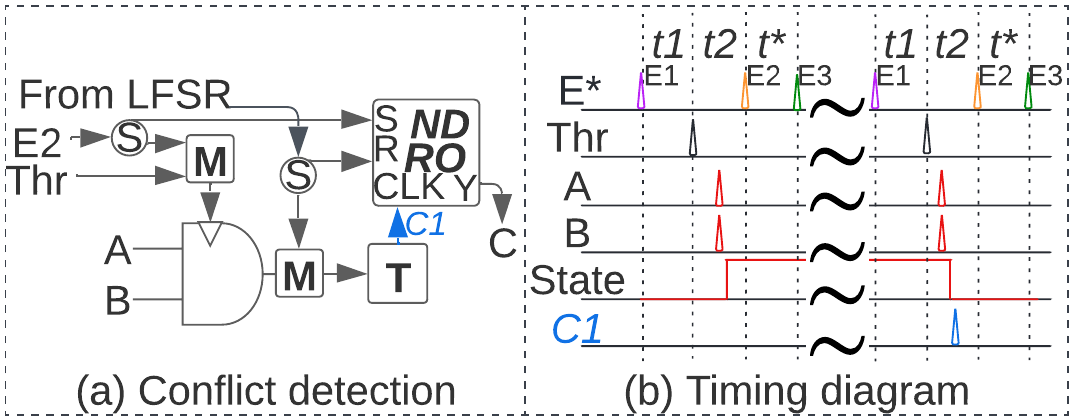}
  \caption{(a) A circuit diagram of the conflict detection logic. $T$ is a \ac{TFF}, $S$ is a splitter cell, and $M$ is a merger cell. (b) A timing diagram that illustrates its operation. $t*$ is the extra control period time slot. $E1$ denotes the beginning of the epoch and thus also the packet's control period, $E2$ the beginning of $t*$, and $E3$ the end of the control period. $E*$ is not a separate signal but rather a notation to illustrate the arrival of both $E1$ and $E2$ in the same waveform line.\label{figure:conflict detection}}
\end{figure}

To aid the round-robin routing logic that follows, we design a conflict detection logic to detect when packets from both inputs request the same output. This happens when the control pulse from inputs $A$ and $B$ arrive both before or both after $Thr$. In case of conflict, this unit changes its internal state and generates a pulse ($C1$) every second conflict. This results in a round-robin resolution of conflicts.

\figurename~\ref{figure:conflict detection} illustrates the conflict detection circuit and an example of operation. The AND gate generates an output pulse if both control pulses arrive between $E1$ and $Thr$ or between $Thr$ and $E2$. If the condition is met, the \ac{TFF} changes its state. $E1$ marks the beginning of an epoch, $E2$ the end of the second time slot in the control period (the beginning of $t*$), and $E3$ the beginning of the data period. In \figurename~\ref{figure:conflict detection}(b), $A$ and $B$ conflict because they request the same output. That does not produce a $C1$ output pulse, but it changes the \ac{TFF}'s internal state. At the next epoch with a conflict, the circuit produces a $C$ pulse and reverts the \ac{TFF}'s internal state.
Note that $E2$, $E3$, and $Thr$ pulses are periodic control signals. Therefore, the conflict detection logic is in principle synchronous because it uses these control signals for timing and to clear its state between epochs. These signals can be either globally distributed or locally derived inside the router from $E1$ using constant delays such as with \acp{JTL}.

\subsubsection{Round-Robin Routing Logic}
\label{section:round robin routing logic}

\begin{figure}
\centering
  \includegraphics[width=\columnwidth]{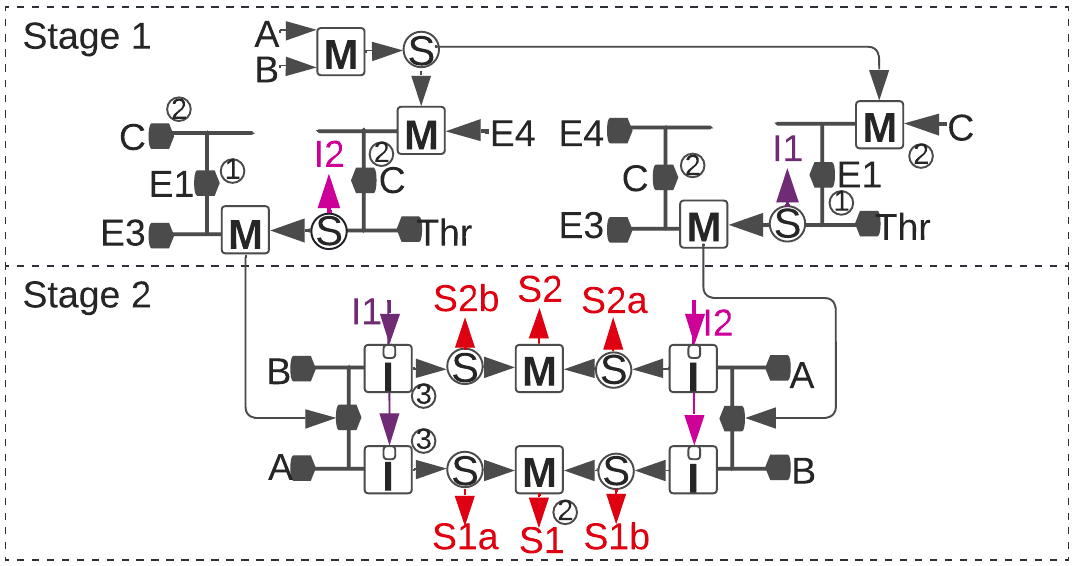}
  \caption{The two-stage round-robin routing logic.\label{figure:round robin routing logic}}
\end{figure}

This logic promotes fairness. When two incoming packets request the same output (a conflict), the routing logic picks a winner packet in a round-robin fashion and deflects the loser packet to the other (not requested) output.
\figurename~\ref{figure:round robin routing logic} shows the round-robin routing logic, which is divided into two stages. The first stage uses the conflict signal ($C$) as input and prepares the second stage. The second stage operates similarly to the fixed-priority logic described above. 

The first stage's primary task is to flip $E1$ and $Thr$ between left and right when it receives a pulse in $C$. This stage is composed of two pairs of \acp{DFF2} facing each other (left and right) through mergers. This stage initially assumes there is no conflict and sets the outer sides of the pairs \textcircled{1} with $E1$. When $E3$ arrives, it propagates through the left \ac{DFF2} at the left pair of the first stage setting the left side of the second stage, which behaves as the fixed priority routing logic of Section~\ref{section:fixed priority routing logic}. In case of conflict, the outer sides of the first stage pairs are reset by $C$ while the inner sides of the pairs are set \textcircled{2}. Also, $E3$ sets the right side of the second stage and $Thr$ sets its left side, achieving a complementary effect. To reduce area, we remove \acp{JJ} for unused \ac{DFF2} outputs (two \acp{JJ} per unused output).

The second stage is built and behaves similarly to the fixed priority routing logic, except that its inputs are $A_{m}$ and $B_{m}$. Also, the second stage's two sides are symmetric given that $E3$ and $Thr$ can arrive at both sides, depending on the first stage.
Furthermore, because control pulses are used by the routing logic to configure the data crossbar, control pulses arrive at the crossbar through the data path before the routing logic has a chance to generate $S1$ or $S2$ to configure the crossbar. This would cause control pulses to be dropped. Therefore, we use splitters in the second stage of the routing logic to generate intermediate signals $S1_a$, $S1_b$, $S2_a$, and $S2_b$; we use these signals to re-generate packet control pulses. Essentially, these four signals codify which input--output combination was selected and which inputs received a packet. For instance, $S1_{a}$ ($S2_{a}$) is asserted when a packet arrived at input $A$ and has been assigned output $A$ ($B$). The new control pulse as a result of these four signals will still be in the same control time slot as the original, thus preserving the packet's encoded destination.

\begin{figure}
\centering
  \includegraphics[width=0.9\columnwidth]{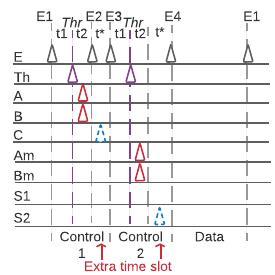}
  \caption{From the control path's perspective, there are two control periods.\label{figure:timing diagram}}
\end{figure}

Using $A_m$ and $B_m$ provides the necessary time for the conflict-detection circuit to identify a conflict and achieve the round-robin deflection algorithm. However, this causes the entire packet to be delayed by a control period, giving the illusion to the control path of two control periods. This is illustrated in \figurename~\ref{figure:timing diagram}.
The first period uses $A$ and $B$ and activates the conflict detection logic while the second uses $A_m$ and $B_m$ and activates the routing logic. The epoch begins when $E1$ arrives and starts the first time slot ($t1$). The second time slot in the first control period ($t2$) starts when $Thr$ first arrives and ends when $E2$ arrives. Between $E2$ and $E3$ is the extra time slot ($t*$). The second control period is from $E3$ until $E4$.
In this example, $A$ and $B$ arrive after $Thr$, indicating that both packets request the same (second) output. Assuming this is the second time this occurs, a conflict pulse ($C$) is generated during the first control period's extra time slot ($t*$). In the shown example, the routing logic resolves this conflict with a $S2$.

\subsubsection{Randomized Round-Robin Routing Logic}
\label{section:randomized round-robin routing logic}

While the round-robin routing logic makes each router divide bandwidth evenly across its inputs by average, under specific circumstances it may cause a packet to be deflected in perpetuity, thus leading to a livelock~\cite{NoCs_bufferless}.
For instance, consider the case where a packet that arrives to a router's input 1 gets deflected because the round robin for the packet's desired output favors input 2; input 2 receives another packet for the same output in the same epoch. Further, lets assume that the deflected packet arrives at an adjacent router at the next epoch and then returns to the router it was deflected at, because that is the shortest path to its final destination. This can occur in the \ac{PaST-NoC} 2D mesh that we describe later. Therefore, the deflected packet arrives to the same router two epochs after it was initially deflected since it has to take two hops. Two epochs after it was deflected, the round robin for the packet's desired output again favors input 2, whereas in this example the packet again arrives to input 1. This process may repeat in perpetuity if input 2 always receives packets that request the same output that conflict with the deflected packet, and the deflected packet returns and re-tries after an even number of epochs. While maintaining these conditions in perpetuity with dynamic traffic is improbable, it is possible.

To prevent livelocks in \ac{PaST-NoC}, we take a probabilistic approach
inspired by the ``chaos'' router~\cite{Chaos_router}. The intuition behind this approach is that we
add randomness in routing decisions such that the repetitious routing patterns that form a livelock
eventually decay. Therefore, any packet has a nonzero probability at every hop of not getting deflected, which means that every packet will eventually get delivered.
This results in a probabilistically livelock-free \ac{NoC}~\cite{Chaos_router}.
To that end, we add randomness to the round-robin conflict detection unit.

Our proposed conflict detection unit has two modes of operation: pure round robin and randomized round robin. To set one mode of operation or the other we use an \ac{NDRO} cell at the output of the conflict detection logic (\figurename~\ref{figure:conflict detection}a).
If an \ac{RSFQ} pulse produced randomly by a \ac{PRNG} such as an \ac{LFSR} does not arrive, the \ac{NDRO} is transparent
because it is set by $E2$ before the conflict detection logic produces a conflict pulse $C1$. Therefore, the router behaves in a pure round-robin fashion. In contrast, in the presence of a random pulse, the \ac{NDRO} is reset. This suppresses the propagation of the conflict pulse $C1$, thus breaking the round-robin sequence. In this case, even if the logic detects a conflict, the routing logic behaves like the fixed-priority routing logic instead of round robin. Finally, the pulse from the \ac{PRNG} is also routed to the \ac{TFF} in order to disrupt the round-robin counter. The insight behind this modification is to allow the conflict detection logic to operate in a round-robin fashion some times, but randomly break the round robin thus disrupting the conditions that cause a livelock other times~\cite{Chaos_router}.

To implement a \ac{PRNG}, we can use existing \ac{RSFQ} multi-bit \acp{PRNG} such as \acp{LFSR}~\cite{LFSR_RSFQ,LFSR_multibit} or other custom \acp{PRNG}~\cite{Random_number_generator}. We can use a single \ac{PRNG} for the entire \ac{NoC} by connecting each bit of a multi-bit \ac{PRNG} to a different \ac{PaST-NoC} router.

While in CMOS a popular solution to livelocks is a form of age-based allocation~\cite{Deflection_age,NoCs_bufferless,Case_bufferless}, this would substantially increase routing logic complexity, the number of \acp{JJ}, and the complexity of a packet's control period. Instead, our proposed solution only adds three cells per router and a \ac{PRNG} per \ac{NoC}. In addition, it does not prolong control periods or other parameters that would degrade throughput or latency.

However small, the area overhead for randomization is not justified for traffic patterns that cannot produce a livelock due to their source--destination pairs or because they pause injecting packets periodically, allowing deflected packets to be delivered.
Therefore, in our evaluations in Section~\ref{section:evaluation}, we use the round-robin routing logic of Section~\ref{section:round robin routing logic} without randomization. However, if livelock prevention is desired, the additional overhead is 24 \acp{JJ} per router and additional \acp{JJ} for the \ac{PRNG} depending on its implementation~\cite{LFSR_RSFQ,LFSR_multibit,Random_number_generator}.

\subsection{Data path}
\label{section:data path}

\begin{figure}
\centering
  \includegraphics[width=\columnwidth]{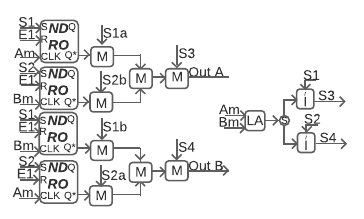}
  \caption{The data crossbar relies on \ac{NDRO} cells to connect each input--output pair. A pulse in $S1$ connects input $A$ to output $A$ and input $B$ to output $B$. A pulse in $S2$ connects input $A$ to output $B$ and input $B$ to output $A$. In parallel, the data crossbar also generates $S3$ and $S4$.\label{figure:data crossbar}}
\end{figure}

\figurename~\ref{figure:data crossbar} shows \ac{PaST-NoC}'s data crossbar that consists of \acp{NDRO} and mergers.
$S1$ sets the first and third \acp{NDRO} from the top, connecting input $A$ to output $A$ and input $B$ to output $B$. $S2$ sets the other two \acp{NDRO}, achieving a complementary effect by connecting input $B$ to output $A$ and input $A$ to output $B$. $A_{m}$ and $B_{m}$ connect to the clock input of the \acp{NDRO}.
A pulse in the clock input of an \ac{NDRO} causes an output pulse if the \ac{NDRO} has received a pulse in its set ($S$) input more recently than a pulse in its reset ($R$) input (i.e., the \ac{NDRO} is ``set''). Therefore, incoming packet pulses through $A_{m}$ and $B_{m}$ are propagated by \acp{NDRO} that are set. Finally, $E1$ resets all \acp{NDRO} to clear the state of all \acp{NDRO} between epochs.

\begin{figure}
\centering
  \includegraphics[width=0.4\columnwidth]{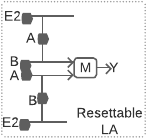}
  \caption{A resettable \ac{LA} gate composed of two \acp{DFF2} and one merger.\label{figure:resettable_la}}
\end{figure}

In parallel, the crossbar also generates $S3$ and $S4$ that supplement $S1_a$, $S1_b$, $S2_a$, and $S2_b$ by reconstructing control pulses in the cases where there are two incoming packets. For this, we design a resettable \ac{LA} gate shown in \figurename~\ref{figure:resettable_la}.

\section{Scaling Up}
\label{section:scaling up}

We can scale up \ac{PaST-NoC} based on the proposed 2$\times$2 router. We first design a 4$\times$4 butterfly \ac{NoC} that we then use as a building block for an 8$\times$8 mesh. These building blocks can then construct a \ac{PaST-NoC} of various scales and topologies.

\subsection{4$\times$4 Butterfly}
\label{section:butterfly}

\begin{figure}
\centering
  \includegraphics[width=0.9\columnwidth]{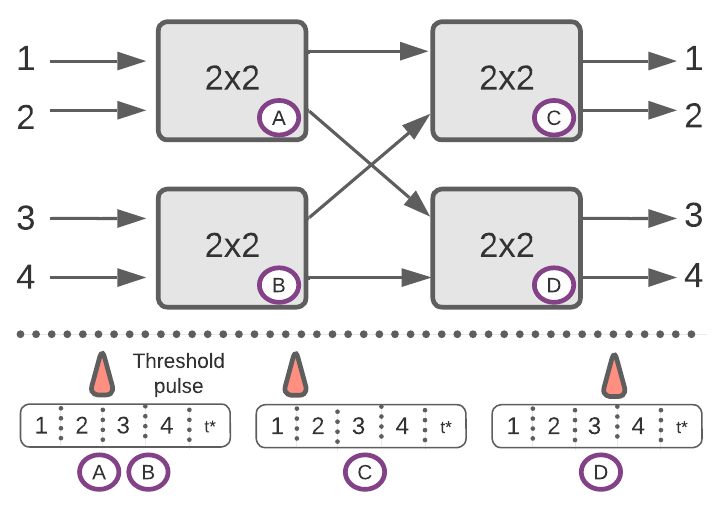}
  \caption{A 4$\times$4 butterfly composed of four 2$\times$2 routers. Depending on its position (\textcircled{A}, \textcircled{B}, \textcircled{C}, or \textcircled{D}) each router has a different timing for its threshold pulse ($Thr$) according to how destinations are mapped to its outputs.\label{figure:butterfly}}
\end{figure}

A 4$\times$4 butterfly is shown in \figurename~\ref{figure:butterfly}.
Each path consists of two hops and there is exactly one path from each source to each destination, which means routing is deterministic.
A butterfly is topologically equivalent to a Banyan topology~\cite{Banyan_network,Banyan2}. 
We design a 4$\times$4 \ac{NoC} with a butterfly instead of scaling up a single router to keep router design complexity low.

To correct for the propagation delay of pulses through the first column routers, $E1$, $E2$, $E3$, and $Thr$ in the routers of the second column are delayed by the propagation delay through a router.
For example, if the propagation delay (latency) of a router is $PD$, all periodic control pulses to routers in the second column are delayed by $PD$.
Otherwise, packets would exit a router in the first column and arrive late at a second-column router, risking that their control and data pulses would appear to be in different time slots.
Likewise, final destinations also delay their epoch start by the propagation delay through two routers ($2 \times PD$), which is the end-to-end propagation delay of this network. The packet latency is $2 \times PD + EpochDuration$ because it includes the time from when a packet starts arriving until it completes.
Note that this discussion applies to any $N$ hop path (in this case $N$ = 2).
An alternative strategy is to prolong time slots to account for propagation delays~\cite{SRNoC}, though that risks degrading throughput.

In this topology, if a packet gets deflected at a router in the first or second column, it will
exit the \ac{NoC} at a destination it did not intend.
Destinations are responsible for re-injecting any packets they receive that are not destined to them~\cite{NoCs_bufferless}.

\figurename~\ref{figure:butterfly} also shows how the timing of the threshold pulse ($Thr$) changes per router, depending on the router's location. This is necessary because each 2 $\times$ 2 router sends to its top output any packet with a control pulse before $Thr$ and others to its bottom output. Therefore, adjusting the timing of $Thr$ at each router is necessary to correctly route packets to final destinations.
From routers \textcircled{A} and \textcircled{B}, in order to reach destinations 1 or 2, a packet has to depart towards the top output that leads to router \textcircled{C}. Likewise, destinations 3 and 4 are reachable from routers \textcircled{A} and \textcircled{B} by their bottom outputs via router \textcircled{D}.
Therefore, for routers \textcircled{A} and \textcircled{B}, the threshold pulse ($Thr$) arrives between control time slots 2 and 3.
However, in router \textcircled{C}, destination 1 is reachable from the top output and destination 2 from the bottom output. Therefore, in router \textcircled{C}, the threshold pulse ($Thr$) arrives after control time slot 1 in order to send packets for destination 1 to the top output and destination 2 to the bottom output. If any packets arrive at router \textcircled{C} for destinations 3 or 4, it is because they were deflected in router \textcircled{A} or \textcircled{B} since neither 3 nor 4 are reachable from router \textcircled{C}. In that case, assuming no second deflection, they will be sent to destination 2 that will re-inject them into the \ac{NoC}.

\subsection{Larger Topologies -- 8$\times$8 2D Mesh}
\label{section:larger topologies}

While we can design a larger butterfly \ac{NoC}, this would increase the hop count as well as the impact of a deflection because if a packet gets deflected at \emph{any} router along its path in a butterfly, it is forced to exit the \ac{NoC} at a destination other than its desired, possibly deflecting other packets along the way.
Adding a stage in the butterfly to increase path diversity lessens this effect, but at a cost of \acp{JJ}.
For instance, an 8$\times$8 butterfly with one extra stage would have 4$\times$ the \acp{JJ} of our 4$\times$4 butterfly.
A single deflection costs $N$ hops in a butterfly, whereas in a mesh, a single deflection costs two hops because a packet can start progressing towards its destination right after it gets deflected once~\cite{NoCs_bufferless}.

\begin{figure}
\centering
  \includegraphics[width=0.9\columnwidth]{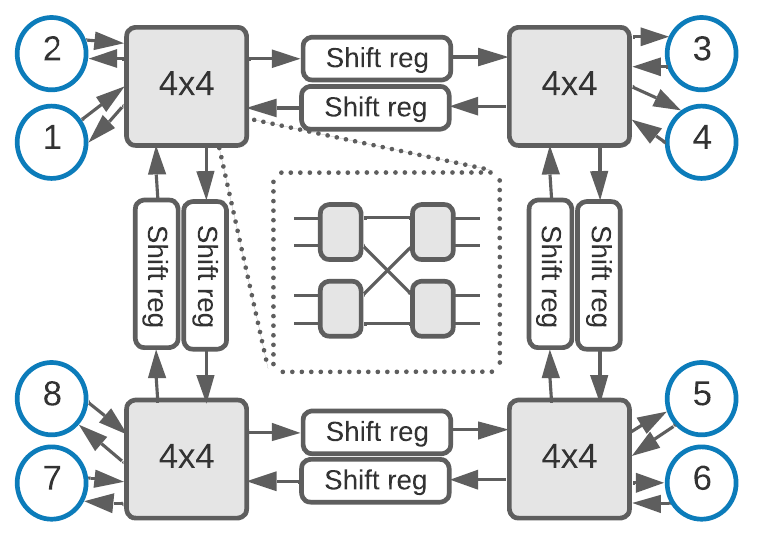}
  \caption{An 8$\times$8 mesh with eight endpoints and four routers. Each router is a 4$\times$4 butterfly and connects to two endpoints (shown as circles).\label{figure:mesh}}
\end{figure}

For this reason and in the interest of exploring different topologies, to scale up \ac{PaST-NoC}, we can treat 4$\times$4 butterfly topologies as individual routers and, combined with our 2$\times$2 routers as additional building blocks, construct a variety of topologies of various scales and dimensions. One such example is a 2D mesh that has four routers arranged in a 2$\times$2 grid and a concentration factor of two. This topology can serve eight endpoints and thus we call it an 8$\times$8 2D mesh, shown in \figurename~\ref{figure:mesh}. In this example, a deflection that occurs inside a butterfly manifests as a deflection at the mesh by a packet exiting a mesh router at an unexpected output.
If that output leads to an endpoint, the deflected packet has to be re-injected into the \ac{NoC}. Otherwise, the packet may proceed farther away from its destination but then has the opportunity to take hops towards its destination~\cite{NoCs_bufferless}.

In this topology, we must delay packets between 4$\times$4 routers such that packets arrive at their next router at the beginning of an epoch.
In contrast to the uni-directional butterfly, because the mesh is bi-directional, we cannot simply delay the periodic signals to some routers. Therefore, we add shift registers between routers. Those shift registers have a total delay of $E - P$ where $E$ is the duration of an epoch and $P$ is the propagation delay through a router (which is a 4$\times$4 butterfly). Each stage has a duration equal to the minimum spacing between data pulses ($SP$) that also defines the period of the clock input to each shift register similarly to Section~\ref{section:router diagram}. Therefore, each shift register has $\frac{E - P}{SP}$ stages. We reduce area overhead by using a flux-based shift register~\cite{Flux_shift_register}.
\section{Evaluation}
\label{section:evaluation}

\subsection{Methodology}
\label{section:methodology}

We use a combination of Verilog, SPICE, a modified version of Booksim that implements deflection flow control~\cite{Booksim,NoCs_bufferless}, and analytical models. We use a bottom--up approach where SPICE validates Verilog that validates Booksim models as scale increases. For SPICE and Verilog models, we build a library with the cells in Table~\ref{table:primitives_summary}.
For SPICE, we use the open-source MIT-LL SFQ5ee 10 kA/cm2 process and WRSPICE, an open source SPICE simulator. WRSPICE simulates our components at the circuit level. Our Verilog models emulate the behavior and approximate delays of each cell observed in SPICE.
We use Verilog models beyond a single 2$\times$2 \ac{PaST-NoC} router after which WRSPICE becomes impractical.
Our WRSPICE decks include testbenches for which we use DC-\ac{RSFQ} cells to take rectangular-like pulses as inputs and output \ac{RSFQ} pulses which resemble a voltage spike. We use a supply and bias voltage of 10 mVs.
In Verilog, in lieu of analog \ac{RSFQ} spikes, we use 5ps long rectangular pulses to represent a pulse.

In both Verilog and WRSPICE, we confirm correct operation by testing all combinations of packets arriving to each input for any of the available destinations, or not arriving at all. We test each case with its own transient simulation as well as periodic repetition to test the round-robin function.

\begin{figure}
    \centering
    \includegraphics[width=\columnwidth]{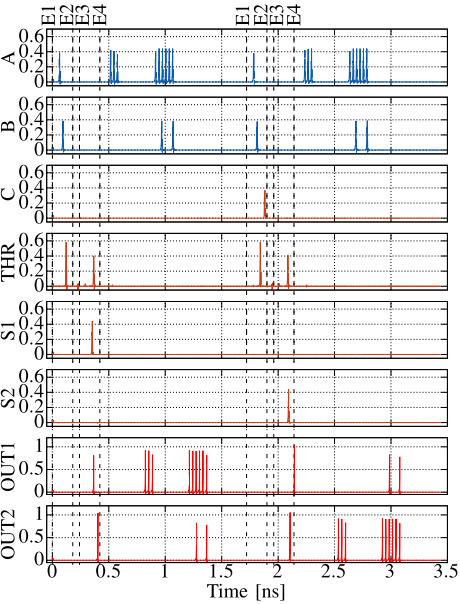}
    \caption{A 2$\times$2 \ac{PaST-NoC} router implemented in WRSPICE. Two epochs are shown. Inputs are shown in blue ($A$ and $B$), internal pulses in orange ($C$, $Thr$, $S1$, and $S2$), and output pulses in red ($Out1$ and $Out2$).\label{figure:wrspice waveform}}
\end{figure}

To assign a data rate to our packets and compare against binary \acp{NoC}, we assume data pulses are encoded in \ac{RL}. To evaluate throughput under high loads, we assume we try to send one pulse per data time slot with each pulse's value determined with a uniform random probability. This mimics a binary \ac{NoC} where packet length equals the size of the information endpoints send per packet. Therefore, we use Equation~\ref{equation:equal} to calculate how many data pulses each packet carries. We report bandwidth in gigabits per second (Gbps) per port after dividing it by the number of \acp{JJ} of each \ac{NoC}, in order to normalize performance per unit area. Reported throughput for \ac{PaST-NoC} is only for packets that arrive to their intended destinations. Both our analytical models and Booksim implementation take into account multiple deflections of a single packet as well as deflections that occur due to previously deflected packets. Also, both models re-inject packets that are ejected from the \ac{NoC} to an undesired destination due to a deflection.

We use detailed analytical models where accurate, in some cases augmented with Booksim simulations, to compare \ac{PaST-NoC} against reported performance numbers from binary \ac{RSFQ} \acp{NoC} due to the difficulty of implementing multiple designs in WRSPICE and in some cases the lack of detailed published information in the architecture or packet format to accurately re-implement binary \acp{NoC}.
To explore the design space, we illustrate \ac{PaST-NoC}'s throughput for different data period durations with a constant control period per packet, determined by the number of destinations. This is roughly equivalent to adjusting the payload size of a binary packet because it changes the control over data ratio.
However, when comparing to prior binary \acp{NoC}, we cannot reliably vary their packet's payload size for our evaluation based
on available information. In fact, in some prior art control information is transmitted via separate wires; those designs are already at the data throughput-optimal point in terms of payload size because data transfer does not pause in order to transmit control information.
Therefore, when evaluating \ac{PaST-NoC}'s throughput for different data period durations, we do not adjust binary \ac{NoC} data throughput.

\subsection{Results}
\label{section:results}

\subsubsection{Timing Parameters}
\label{section:timing parameters}

To preserve pulse count and integrity the minimum spacing of data pulses is 15ps.
This guarantees that pulses will not overlap in any cell in the datapath and \ac{NDRO} cells function correctly.
Therefore, a data period of 300ps can fit at most $\frac{300}{15} = 20$ data pulses. Additionally, the minimum duration of a time slot in the control period is 60ps. Therefore, in a \ac{PaST-NoC} with two destinations, the control period lasts 180ps (two time slots plus one extra).
Our WRSPICE circuit models include wires as inductors and parasitic inductance.

\subsubsection{2$\times$2 Router}
\label{section:2x2 router}

\figurename~\ref{figure:wrspice waveform} shows the round-robin deflection functionality of a 2$\times$2 \ac{PaST-NoC} router in the case when two packets request the same output for two consecutive epochs. Packets arrive in each input and request output 1 since both packet's control pulses arrive before $Thr$. In the first epoch, the packet from input $A$ is chosen as shown by a pulse in $S1$. In the second epoch, $C$ triggers and causes an $S2$ instead. Thus, the packet from input $B$ is chosen. In each epoch, the non-chosen packet is deflected to output 2. Data pulses of each packet follow the packet's control pulse.

\begin{figure}
\centering
  \includegraphics[width=\columnwidth]{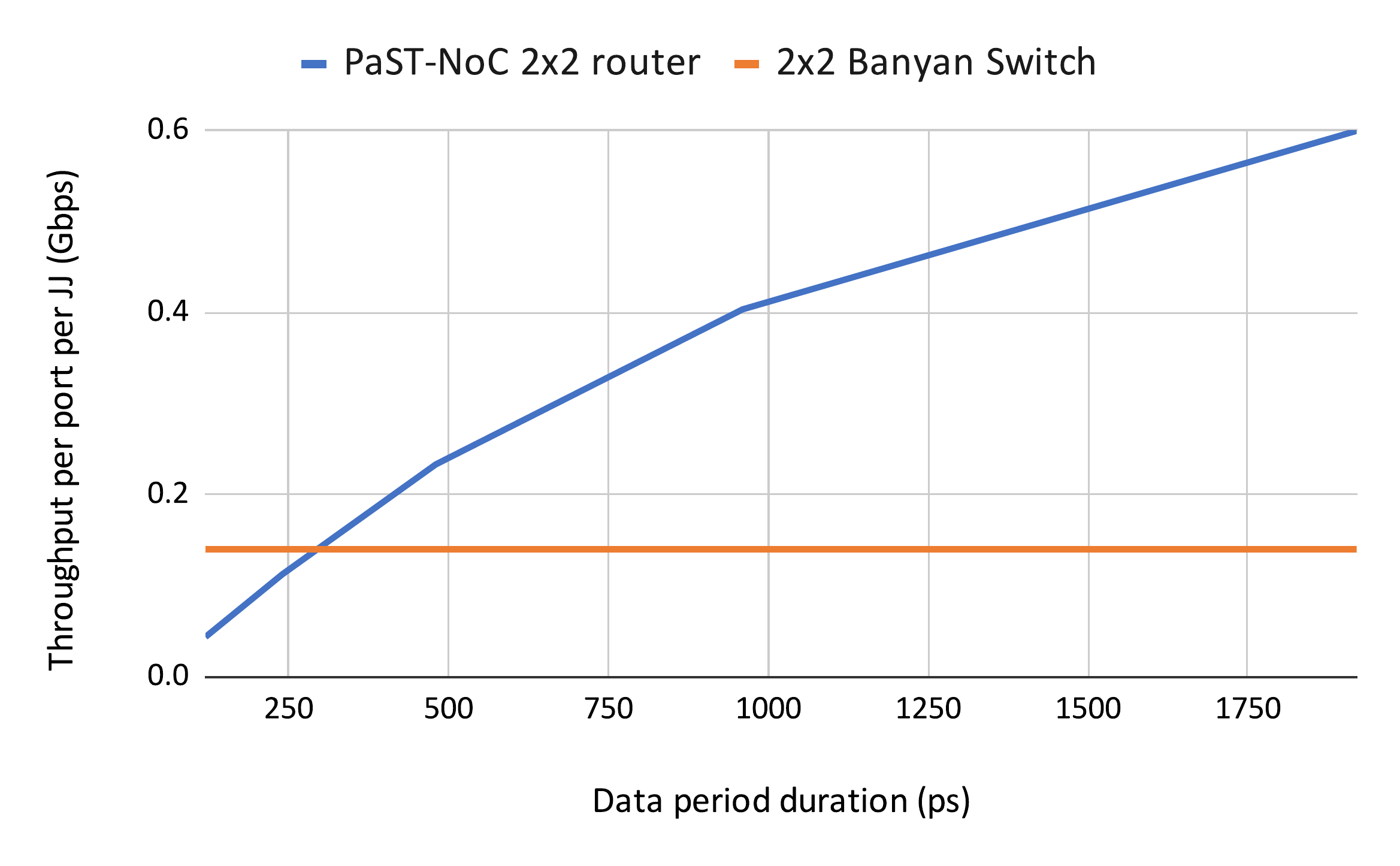}
  \caption{Throughput per port per \ac{JJ} for a 2$\times$2 binary switch~\cite{Banyan2} and for a 2$\times$2 \ac{PaST-NoC} router, as a function of the duration of a packet's data period.\label{figure:2x2 curve}}
\end{figure}

\begin{table}
\begin{center}
\caption{2$\times$2 \ac{PaST-NoC} router \ac{JJ} count and delay breakdown.\label{table:router breakdown}}
\begin{tabular}{|c|c|c|}
\hline
\rowcolor{LightCyan}
Module & Number of \acp{JJ} & Prop. delay (ps)\\
\hline
\hline
Conflict detection & 27 & 40.95\\
\hline
Routing logic stage 1 & 87 & 50\\
\hline
Routing logic stage 2 & 91 & 41.06\\
\hline
Data crossbar & 89 & 33.9\\
\hline
Resettable \ac{LA} & 34 & 28.95\\
\hline
Shift register & 44 & 162.17\\
\hline
Miscellaneous & 109 & -- \\
\hline
\hline
Summary & Total \textbf{481} & In to out \textbf{213.41}\\
\hline
\end{tabular}
\end{center}
\end{table}

Table~\ref{table:router breakdown} provides a breakdown of the number of \acp{JJ} and propagation delay for different modules of a \ac{PaST-NoC} router. We measure delay from the time the last input arrives to each module, until it produces its last output. We report the worst case delay (for different input patterns) for each module. ``In to out'' delay is the propagation delay through the router, from a packet first entering an input to starting to depart from an output.
``Miscellaneous'' refers to cells that are not explicitly assigned to a module, such as splitters and mergers that implement fanin and fanout.
We notice a variability in the datapath of about $\pm$10ps due to the flux-based shift register. This variability can be drastically reduced with \ac{DFF}-based shift registers but at a \ac{JJ} cost. Lower variability can allow shorter control periods.

\begin{figure}
    \centering
    \includegraphics[width=\columnwidth]{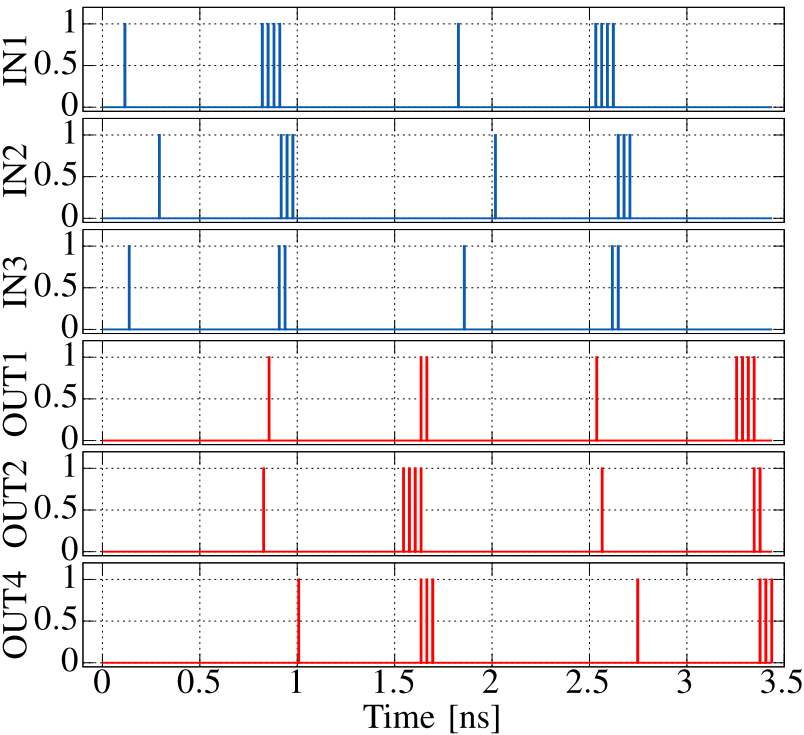}
    \caption{The 4$\times$4 \ac{PaST-NoC} butterfly of \figurename~\ref{figure:butterfly} implemented in Verilog. Two epochs are shown. Input pulses are shown in blue ($In1$, $In2$, $In3$) and output pulses in red ($Out1$, $Out2$, $Out3$).\label{figure:verilog butterfly}}
\end{figure}

In addition, we measure the static power of one \ac{PaST-NoC} 2$\times$2 router to be 665.56$\mu$W, by measuring current draw. Worst case dynamic power, measured with a high frequency of pulses at a case where two incoming packets conflict is 195nW. Static power is much higher than dynamic power because our experiments are based on \ac{RSFQ} technology
that uses on-chip bias resistors that constantly consume power~\cite{Superconducting_accelerators,RSFQ}.
To compare, we estimate the static power of the binary 2$\times$2 router of~\cite{Banyan2} from its schematic using the measured static power draw from individual cells in our library.
The calculated static power is 1.4mW (approximately 2$\times$ that of a \ac{PaST-NoC} router), including the clock tree that by itself consumes 15\% more static power than a \ac{PaST-NoC} router.

\figurename~\ref{figure:2x2 curve} shows an analytical study of how average throughput per port per \ac{JJ} varies in a 2$\times$2 \ac{PaST-NoC} router as a function of the duration of the data period. This analysis is for \ac{UR} traffic and includes the probability for a packet to be deflected, which is 25\% for \ac{UR} traffic.
Doubling the data period does not double throughput, especially for longer data periods, because of the analysis of Equation~\ref{equation:equal}.
Note that for a single hop, there is no appreciable sustained data rate difference between a blocking \ac{NoC} and \ac{PaST-NoC} that re-injects deflected packets.
For comparison, \figurename~\ref{figure:2x2 curve} also includes calculated performance metrics for a 60-gate, 1184-JJ, 40GHz 2$\times$2 binary switch for a Banyan topology~\cite{Banyan2}. As shown, the crossover point is approximately a data period of 300ps.

\subsubsection{4$\times$4 Butterfly}
\label{section:4x4 butterfly}

To implement the 4$\times$4 butterfly of Section~\ref{section:4x4 butterfly} that has 1924 \acp{JJ}, we increase the control period by two time slots for a total of 300ps.
We demonstrate the functionality using Verilog in \figurename~\ref{figure:verilog butterfly}.
As shown, packets from inputs 1 and 3 are both destined for destination 2 ($Out2$). Therefore, they conflict in router \textcircled{C} in the second column of \figurename~\ref{figure:butterfly}. In this example, input 1's packet gets selected. Input 3's packet gets deflected to destination 1. Also, in parallel, a packet from input 2 exits at destination 4 without causing any conflicts. In the next epoch, traffic repeats but router \textcircled{C} now deflects input 1's packet to destination 1.

We then compare against two state of the art 4$\times$4 binary \acp{NoC}: (i) a 4$\times$4 Banyan topology with 4300 \acp{JJ}~\cite{Banyan_network} and (ii) a 4$\times$4 crossbar with 4316 \acp{JJ}~\cite{Crossbar3}. For each, we calculate performance metrics based on reported values. We also compare against a 4$\times$4 \ac{SRNoC} router with 528 \acp{JJ} and 64 time slots per connection window~\cite{SRNoC}.

For \ac{PaST-NoC}, we calculate average deflection rates and subsequently average throughput for the best case, \ac{UR}, and worst case traffic. All other traffic patterns fall within this range. Best case traffic is one with no deflections. \ac{UR} has a 25\% deflection probability at every hop. Worst case traffic is one where in every first-stage router packets ask for the same output; therefore, packets have a 50\% deflection probability at the first hop and then 25\% at the second hop, since deflections help load balance~\cite{NoCs_bufferless}.
For \ac{SRNoC}, we compare the best case of \ac{PaST-NoC} against the best case of \ac{SRNoC} (\ac{UR}) and the worst case of \ac{PaST-NoC} against the worst case of \ac{SRNoC}, which is a source sending at full rate to a single destination.
All traffic patterns we evaluate are admissible, i.e., no destination receives more than 100\% of traffic.

\begin{figure}
\centering
  \includegraphics[width=\columnwidth]{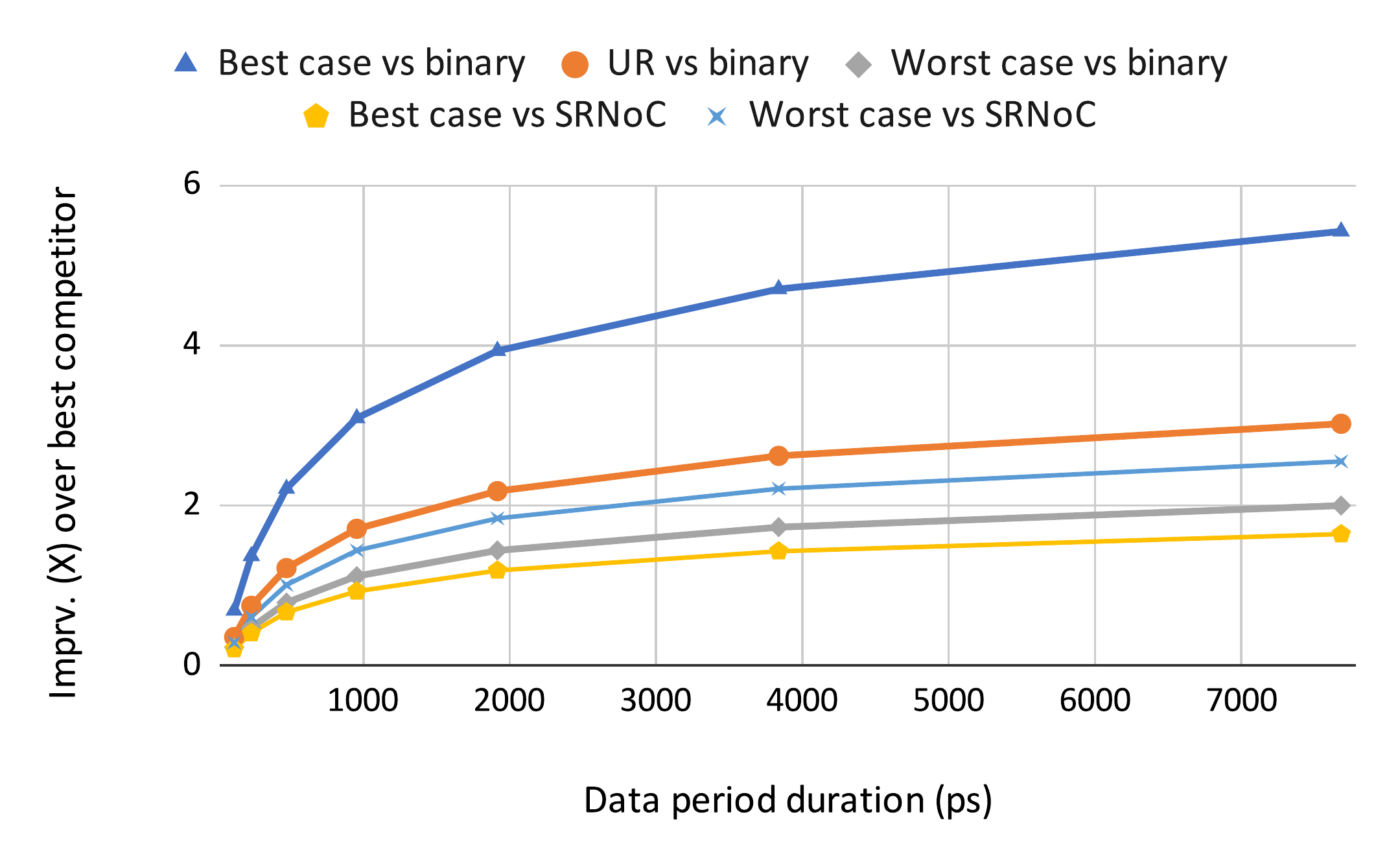}
  \caption{Improvement factor ($\times$) in throughput per port per \ac{JJ} of a \ac{PaST-NoC} 4$\times$4 butterfly as a function of a packet's data period duration. Results are shown separately against \ac{SRNoC} and the best of our two binary competitors.\label{figure:4x4 curve}}
\end{figure}

\figurename~\ref{figure:4x4 curve} shows the improvement factor ($\times$) in throughput per port per \ac{JJ} of our 4$\times$4 \ac{PaST-NoC} butterfly
when compared to the best of the two aforementioned state of the art \ac{RSFQ} binary \acp{NoC} with 4316 and 4781 JJs respectively, and separately when compared against \ac{SRNoC}.
An improvement factor of less than 1, equal to 1, or over 1 means that \ac{PaST-NoC} has a lower, equal, or greater throughput per \ac{JJ} compared to the best competitor, respectively.
For the binary comparison, best case traffic has a crossover point of 450ps for the data period duration, \ac{UR} has 930ps, and worst case traffic 1890ps.
When compared to \ac{SRNoC}, \ac{PaST-NoC}'s worst case is more favorable because \ac{SRNoC}'s worst case only uses a quarter of available bandwidth whereas \ac{PaST-NoC} is a packet-switched \ac{NoC} and thus adapts better to unbalanced traffic.
For best traffic, the crossover point against \ac{SRNoC} is 960ps and for worst case 465ps.

\subsubsection{8$\times$8 Mesh}
\label{section:8x8 mesh}

\begin{figure}
    \centering
    \includegraphics[width=\columnwidth]{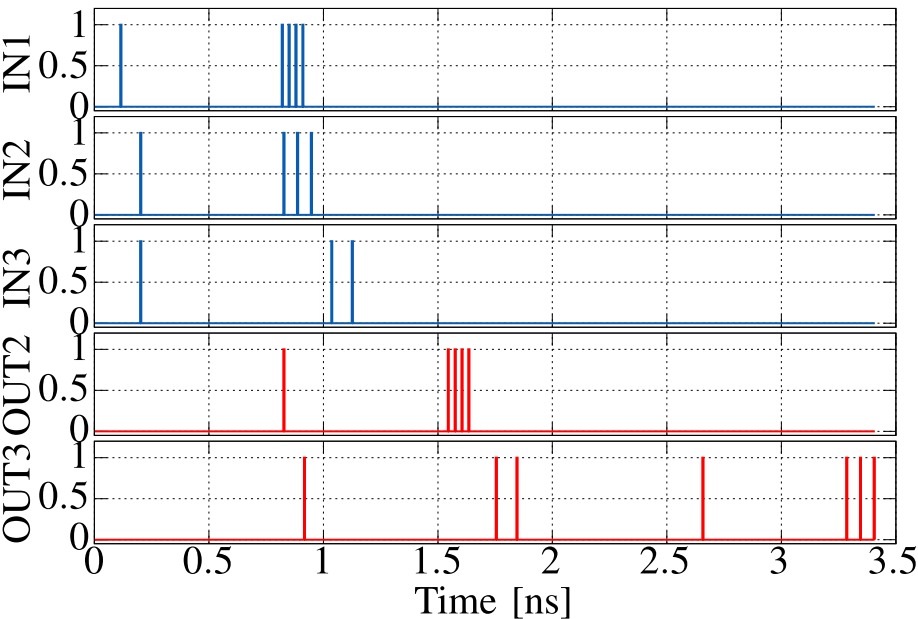}
    \caption{Verilog simulation of the 8$\times$8 2D mesh of \figurename~\ref{figure:mesh}. Input pulses are shown in blue ($In1$, $In2$, $In3$) and output pulses in red ($Out2$, $Out3$).\label{figure:verilog mesh}}
\end{figure}

Here we evaluate our 8$\times$8 2D mesh of Section~\ref{section:larger topologies} that has 7912 \acp{JJ}.
\figurename~\ref{figure:verilog mesh} demonstrates its functionality in a case where three packets traverse the mesh. The packet from input 1 is destined to destination 2 ($Out2$). The packets from inputs 2 and 3 both are destined to destination 3. The packet from input 2 has a greater hop count because it travels from the top left 4$\times$4 router to the top right. Therefore, the two packets do not meet each other and exit at destination 3 in sequence with an empty epoch in between. This matches the behavior of a binary \ac{NoC} with deflection flow control.

To evaluate throughput, we analytically calculate the total number of \acp{JJ} including the inter-router shift registers. We then calculate the maximum throughput per port per \ac{JJ} for \ac{PaST-NoC}, assuming no deflections. To account for deflections, we use Booksim to simulate five synthetic traffic patterns using \ac{DOR}. In addition, we construct a synthetic traffic pattern that mimics temporal and spatial imbalance found in benchmarks. For each traffic pattern, we report the worst case (among all endpoints) percentage of the maximum throughput that an endpoint can achieve. We then use these percentages to scale down the throughput per port per \ac{JJ} for \ac{PaST-NoC}, compared to the maximum. In practice, benchmarks typically do not constantly inject at maximum throughput, thus their throughput is less penalized by deflections than what we report in our results for \ac{PaST-NoC}. Booksim simulations use single-flit packets, which mimics \ac{PaST-NoC}'s packets of the same length that propagate as a single unit. In addition, we scale up the control period to 540ps to address eight destinations.

\begin{figure}
\centering
  \includegraphics[width=\columnwidth]{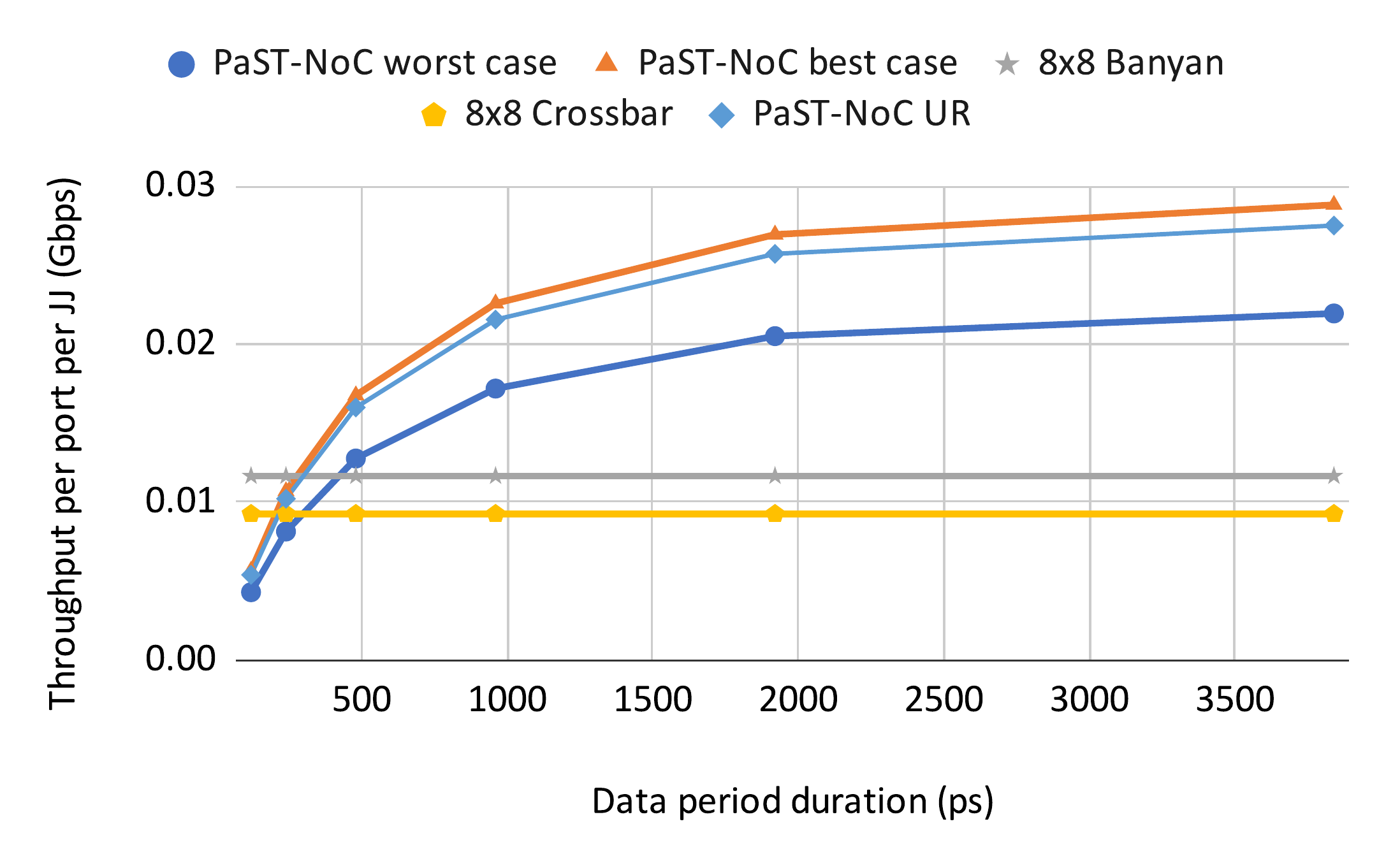}
  \caption{Throughput per port per \ac{JJ} for the 8$\times$8 \ac{PaST-NoC} mesh of \figurename~\ref{figure:mesh} as a function of the control period duration, an analytical model of an 8$\times$8 Banyan network constructed from 2$\times$2 switches~\cite{Banyan2}, and an analytical model of an 8$\times$8 crossbar constructed from four 4$\times$4 crossbars~\cite{Crossbar3}.\label{figure:mesh curve}}
\end{figure}

For comparison, we analytically calculate throughput per port per \ac{JJ} for (i) an 8$\times$8 binary Banyan \ac{NoC} constructed from twelve 2$\times$2 switches~\cite{Banyan2} and (ii) an 8$\times$8 binary \ac{NoC} constructed out of four 4$\times$4 crossbars~\cite{Crossbar3} (the area of an N$\times$N fully-connected crossbar increases quadratically with $N$). For both competitors, the throughput per port of all admissible traffic patterns equals that of ideal traffic. However, throughput per port per \ac{JJ} still is affected by the number of \acp{JJ}.

\figurename~\ref{figure:mesh curve} shows that \ac{PaST-NoC} outperforms both competitors for data periods longer than approximately 360ps for worst case traffic and 255ps for \ac{UR} and best case traffic. The inter-router shift registers (\figurename~\ref{figure:mesh}) have 42 to 290 stages each and collectively contribute 5\% to 30\% of the overall \ac{PaST-NoC} \ac{JJ} count, depending on the data period duration.

\subsubsection{Scalability}
\label{section:scalability}

The scales we study already approach the practical \ac{JJ} count limits given \ac{RSFQ}'s limited device density~\cite{Superconducting_accelerators, RSFQ_scalability}, especially for chips that also contain memory elements.
However, scaling up to more destinations favors binary \acp{NoC} because \ac{PaST-NoC}'s control period scales linearly whereas a binary packet's destination field scales logarithmically.
Using analytical evaluations, with load balanced traffic \ac{PaST-NoC} outperforms all binary \ac{RSFQ} \acp{NoC} for up to a few hundred endpoints. Even under worst-case traffic and throughput assumptions, \ac{PaST-NoC} remains favorable for a few tens of endpoints for long data periods.
Deflection flow control has been shown to scale efficiently well within the scales we study~\cite{Case_bufferless,NoCs_bufferless}.

\subsubsection{32$\times$32 Evaluation}
\label{section:32x32 evaluation}

\begin{figure}
    \centering
    \includegraphics[width=\columnwidth]{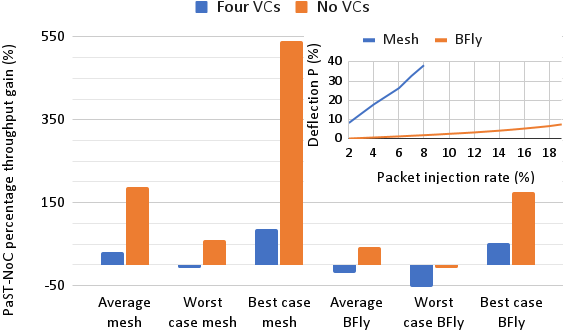}
    \caption{\ac{PaST-NoC} throughput improvements as a percentage (\%) for a 32$\times$32 concentrated 2D mesh with four endpoints per 8$\times$8 router, and a five-stage 32$\times$32 butterfly (BFly) with 2$\times$2 routers. The embedded graph (top right) shows in the Y axis the probability (\%) for a packet to be deflected in an 8$\times$8 router in the mesh network or a $2\times$2 router in the butterfly network, by average across our traffic patterns. The X axis shows the packet injection rate (\%) until the first pattern saturates in each of the two \ac{PaST-NoC} topologies.\label{figure:booksim results}}
\end{figure}

We scale up our throughput evaluation to 32 inputs and 32 outputs using a modified version of Booksim that simulates deflection flow control~\cite{NoCs_bufferless}. We simulate two topologies: (i) a 2D concentrated mesh with four terminals per router (routers are 8$\times$8), and (ii) a five-stage butterfly with 2$\times$2 routers. For the mesh, we design an 8$\times$8 \ac{PaST-NoC} router from a three-stage butterfly. We compare \ac{PaST-NoC} against a buffered \ac{NoC} with credit-based flow control and either no \acp{VC} (one buffer per input) or four \acp{VC} to show a higher-performing variation. The no \ac{VC} case matches previously-demonstrated \ac{RSFQ} binary \acp{NoC}. The butterfly uses destination tag single-path routing while the no \ac{VC} mesh \ac{DOR}. For four \acp{VC} in the mesh, we keep the highest throughput among deterministic \ac{DOR}, oblivious XY-YX routing, and minimal adaptive~\cite{Booksim}. We adjust the throughput per port for \ac{PaST-NoC} based on our average results for 1000ps data periods from \figurename~\ref{figure:mesh curve} and \figurename~\ref{figure:4x4 curve}, as well as the control period scaling characteristics of Section~\ref{section:scalability}. We use cycle times that are based on simple \ac{RSFQ} \acp{NoC} without \acp{VC} for all buffered cases in Booksim, even though \acp{VC} increase complexity thus prolonging the critical path. We also assume all buffered routers have one pipeline stage and ignore additional area to implement \acp{VC}. All these assumptions favor the buffered case.
\figurename~\ref{figure:booksim results} shows the throughput for the \ac{PaST-NoC} average, best, and worst case over five synthetic traffic patterns: \ac{UR}, tornado, bitcomp, shuffle, and transpose~\cite{Booksim}.

As shown, \ac{PaST-NoC} outperforms the no \ac{VC} buffered case by an average 188\% for the mesh and 43\% for the butterfly. The best case for \ac{PaST-NoC} for the mesh is for the tornado traffic pattern where deflection acts to load balance traffic better than any of our three routing policies in the buffered case. For the butterfly, the worst case for \ac{PaST-NoC} is \ac{UR} traffic where transient imbalance can cause a deflection in any of the five hops. In that case, because the butterfly is not bi-directional and has no path diversity, even a single deflection forces the packet to fully exit and re-enter the network; in the mesh a single deflection just costs two hops (Section~\ref{section:larger topologies}). \ac{PaST-NoC} outperforms the four \ac{VC} case in half the metrics, and the no \ac{VC} case in all experiments except for one traffic pattern in the butterfly where \ac{PaST-NoC} shows a 7\% penalty.

\figurename~\ref{figure:booksim results} also shows the per-packet \ac{PaST-NoC} deflection probability. Furthermore, we notice that even though latency increases more rapidly with injection rate than buffered \acp{NoC}~\cite{NoCs_bufferless,Case_bufferless}, the higher bandwidth of \ac{PaST-NoC} results in lower network load for a constant bits per second of injection rate. Therefore, in test cases where \ac{PaST-NoC} outperforms buffered \acp{NoC}, the latency impact of deflection is below 2\% for injection rates that do not saturate the buffered \acp{NoC}.

\subsection{CMOS Comparison}
\label{section:cmos comparison}

Using \ac{ERSFQ}~\cite{mukhanov_tasc_2011} we can eliminate static power consumption. We can also account for the power associated with cooling by multiplying the circuit power by 400$\times$~\cite{ishida_micro_2020}. In this case, based on our \ac{JJ} count and power results from Section~\ref{section:2x2 router}, dynamic power for a 2$\times$2 \ac{PaST-NoC} router becomes a few $\mu$W. Specifically, including the cooling overhead, our 195nW measurement for worst-case dynamic power becomes $195 nW \times 400 = 78 \mu$W. In an ``average'' case where only one packet traverses the 2$\times$2 router, dynamic power is approximately half of that. Even with a 50\% dynamic power penalty to account for \ac{ERSFQ}'s multiplicative power overhead relative to pure logic components~\cite{ERSFQ_penalty}, worst-case total power for one 2$\times$2 \ac{PaST-NoC} router remains below 150 $\mu$W. This number is a few orders of magnitude lower than the mW typically reported in modern CMOS \acp{NoC}. For instance, DDRNoC and ArSMART report one to eight W total for an $8 \times 8$ 2D mesh depending on load~\cite{DDR_NoC,ArSMART_NoC}. This results in approximately 10-150 mW (67$\times$ to 1000$\times$  higher) per router for both CMOS \acp{NoC}.

Comparing throughput between CMOS \acp{NoC} and \ac{PaST-NoC} depends on multiple parameters. Under low enough injected load, all \acp{NoC} satisfy injected traffic thus provide the same throughput. Therefore, \acp{PaST-NoC} provides higher throughput per power because it reduces power as explained above assuming \ac{ERSFQ}. Using as an example our $32 \times 32$ 2D mesh of Section~\ref{section:32x32 evaluation} under \ac{UR} traffic, DDRNoC~\cite{DDR_NoC} has an approximately 10$\times$ higher maximum throughput (Gbps) per port than \ac{PaST-NoC}, assuming 16 time slots per packet and the analysis of Equation~\ref{equation:equal}. However, even after taking into account the best case router power numbers for CMOS \acp{NoC}, \ac{PaST-NoC} provides $6.7 \times$ higher throughput per unit power. This analysis only takes into account router power, but including wire power will make the final result more favorable for \ac{PaST-NoC} due to the high power contribution of wires in CMOS \acp{NoC}~\cite{NoCs_bufferless}, which is significantly lower in \ac{RSFQ}~\cite{Interconnect_routing}.

Finally, if we use \ac{JJ} area models for a modern manufacturing process, a 2$\times$2 \ac{PaST-NoC} router occupies approximately a few hundredths of $mm^2$. Even with a 3$\times$ area penalty for non-\ac{JJ} components, routing, and layout, \ac{PaST-NoC} occupied area is on par with modern CMOS \acp{NoC} despite the stark device density difference.

\section{Discussion and Future Work}
\label{section:discussion and future work}

High timing variability may necessitate more conservative pulse spacing or prolonging control periods and time slots. In the future, we can investigate variable-length epochs or coalescing packets to support variable-size packets.
We can also investigate zero-payload packets; this can support the creation of circuits or channel reservations. While we can design larger-scale versions of a butterfly and a mesh, \ac{PaST-NoC} can also construct numerous other topologies such as a fat tree and a torus by using the same building blocks. Finally, shift registers within and across routers are just retiming elements, similar to pipeline \acp{DFF} in CMOS; packets do not stall in shift registers or elsewhere in \ac{PaST-NoC}, and thus cannot deadlock.
\section{Conclusion}
\label{section:conclusion}

We propose \ac{PaST-NoC}, a packet-switched superconducting \ac{NoC} with deflection flow control that operates its control path using \ac{RL}, addressing the stringent area constraints of superconducting technology. \ac{PaST-NoC} packets carry data pulses that receivers can interpret as pulse trains, \ac{RL}, serialized binary, or other formats. \ac{PaST-NoC} scales to arbitrary topologies based on 2$\times$2 routers and 4$\times$4 butterflies. \ac{PaST-NoC} outperforms binary \ac{RSFQ} \acp{NoC} by as much as over $5\times$.

\section*{Acknowledgments}
This work was supported by the Director, Office of Science, of the U.S. Department of Energy under Contract No. DE- AC02-05CH11231.

\bibliographystyle{IEEEtran}
\bibliography{bibliography.bib}

\vfill

\end{document}